\documentclass{PoS}

\usepackage{amsmath}
\usepackage{dsfont}

\title{\vspace*{-2.5cm}
{\hfill \texttt{\footnotesize Preprint CERN-PH-TH-2014-227}}
\vfill
Condensation phenomena in two-flavor scalar QED\\
at finite chemical potential}

\ShortTitle{Condensation phenomena in scalar QED}

\author{\speaker{Alexander Schmidt}
  \thanks{As part of this work was done together with Philippe de Forcrand during a research stay at the European Organization for Nuclear Research, the author would like to thank CERN and especially Philippe for their hospitality.}
  \thanks{Alexander Schmidt is funded by the FWF DK W1203 "Hadrons in Vacuum, Nuclei and Stars". 
    This work is partly supported by DFG TR55, "Hadron Properties from Lattice QCD"
    and by the Austrian Science Fund FWF Grant. Nr. I 1452-N27.}\\
  Institut f\"ur Physik, Karl-Franzens Universit\"at Graz, Austria\\
  E-mail: \email{alex@treefish.org}}

\author{Philippe de Forcrand\\
        Institut f\"ur Theoretische Physik, ETH Z\"urich, Switzerland, and\\
        Physics Department, TH Unit, CERN, Switzerland\\
        E-mail: \email{forcrand@phys.ethz.ch}}

\author{Christof Gattringer\\
        Institut f\"ur Physik, Karl-Franzens Universit\"at Graz, Austria\\
        E-mail: \email{christof.gattringer@uni-graz.at}}

\abstract{We study condensation in two-flavored, scalar QED with non-degenerate masses at finite chemical potential. The conventional formulation of the theory has a sign problem at finite density which can be solved using an exact reformulation of the theory in terms of dual variables. We perform a Monte Carlo simulation in the dual representation and observe a condensation at a critical chemical potential $\mu_c$.

After determining the low-energy spectrum of the theory we try to establish a connection between $\mu_c$ and the mass of the lightest excitation of the system, which are naively expected to be equal.
It turns out, however, that the relation of the critical chemical potential to the mass spectrum in this case is non-trivial: Taking into account the form of the condensate and making some simplifying assumptions we suggest an adequate explanation which is supported by numerical results.
}

\FullConference{The 32nd International Symposium on Lattice Field Theory,\\
		23-28 June, 2014\\
		Columbia University New York, NY}

\begin{document}

\section{Introduction}
We study scalar QED with two flavors of complex, massive fields $\phi$ and $\chi$ on the lattice, where the masses of the two fields are chosen 
differently. For each flavor a finite chemical potential coupling to the conserved $U(1)$ charge is introduced to investigate condensation. In the 
conventional representation the theory has a sign problem at finite density, which we can overcome by rewriting the partition sum in terms of dual 
variables. For the theory we are studying here, a detailed treatment of the dual reformulation and the description of suitable simulation algorithms 
can be found in \cite{thisone}. Similar  techniques can be used to solve the sign problem of other theories, see, e.g., \cite{others}. 

In this study, we are mainly interested in the connection of the condensation thresholds to the masses of the lowest-lying excitations of the system. This is motivated by the question whether one can see 
sequential condensation when the chemical potential is coupled to charges carried by more than one
field, or having different quantum numbers. A particularly intriguing instance of such a situation is that of QCD with isospin
chemical potential: When the chemical potential is increased, can one observe sequential condensation of the
charged pion, then the charged rho, et cetera? We study this situation in a simple two-flavor toy model. For a related study
in a theory without sign problem see \cite{g2}.

Against our expectations we only find a single condensation threshold, where the original motivation for this study, as already pointed out, was to possibly find a sequence of thresholds. Nonetheless, it turns out that understanding the non-trivial relation of the mass spectrum to the determined critical chemical potential is interesting in its own right and we will present an adequate explanation for the obtained results.

\section{Conventional lattice action}
\label{secconvaction}
We write the action of two-flavor scalar QED with non-degenerate quark masses
as a sum,
\begin{equation}
  S[U,\phi,\chi] \; = \; S_U[U] \; + \; S_\phi[\phi,U] \; + \; S_\chi[\chi,U] \quad .
\end{equation}
Here $S_U$ is the pure gauge action, while $S_\phi$ and $S_\chi$ denote the actions of
the matter fields $\phi$ and $\chi$ repectively.
The gauge part $S_U$ of the action is given by the usual Wilson plaquette action,
\begin{equation}
  S_U[U] \; = \; - \beta \sum_x \sum_{\sigma < \tau} 
  \mbox{Re} \; U_{x,\sigma} U_{x+\widehat{\sigma}, \tau}
  U_{x+\widehat{\tau},\sigma}^\star U_{x,\tau}^\star \quad ,
\end{equation}
where $\beta$ is the inverse gauge coupling, and the sum runs over the real part of all unique plaquettes
$\mbox{Re} \; U_{x,\sigma} U_{x+\widehat{\sigma}, \tau}U_{x+\widehat{\tau},\sigma}^\star U_{x,\tau}^\star$, with $U_{x,\sigma}$
being a link variable at lattice site $x$ in direction $\sigma$. The gauge fields $U_{x,\sigma}$ are elements of $U(1)$. Here $x$ specifies lattice sites on a
4-dimensional space-time lattice with extent $N_s$ in the three spatial directions, and $N_t$ in the Euclidean
time direction.

The matter part of the action involving the complex valued fields $\phi \in \mathds{C}$ is given by
\begin{align}
  \label{phiaction}
  S_\phi   
  \; = \; \sum_x \!\Big(  M_\phi^2 \, |\phi_x|^2  + \lambda_\phi |\phi_x|^4  -
  \sum_{\nu = 1}^4 \!
  \big[  e^{-\mu_\phi  \delta_{\nu, 4} } \, \phi_x^\star \, U_{x,\nu} \,\phi_{x+\widehat{\nu}} 
  \, + \, 
  e^{+\mu_\phi \delta_{\nu, 4}} \, \phi_x^\star \, 
  U_{x-\widehat{\nu}, \nu}^\star \, \phi_{x-\widehat{\nu}}  \big] \!  \Big) \quad .
\end{align}
We use the abbreviation $M_\phi^2=8+m_\phi^2$, with $m_\phi^2$ being the bare mass parameter
of the field $\phi$. The second parameter $\lambda_\phi$ is the coupling for a quartic self-interaction.
Note that a chemical potential $\mu_\phi$ is introduced in the usual way by adding a term proportional to
$\mu_\phi Q$ to the Hamiltonian of the system, where $Q$ is the conserved charge, connected to the $U(1)$ symmetry of
the field $\phi$. As usual, the chemical potential $\mu_\phi$ induces an imbalance between hopping terms in 
positive and negative imaginary time direction, as can be seen from equation (\ref{phiaction}) and thus the action $S_\Phi$ 
is complex for $\mu_\phi \neq 0$. The sum 
$\sum_x$ in (\ref{phiaction}) runs over all lattice sites $x$ and for the hopping terms there is an
additional sum over all directions $\nu=1,...,4$.

The action $S_\chi$ for the second flavor, is almost identical to $S_\phi$,
\begin{align}
  S_\chi
  &= \sum_x \!\Big(  M_\chi^2 \, |\chi_x|^2  + \lambda_\chi |\chi_x|^4  -
  \sum_{\nu = 1}^4 \!
  \big[  e^{-\mu_\chi  \delta_{\nu, 4} } \, \chi_x^\star \, U_{x,\nu}^\star \,\chi_{x+\widehat{\nu}} 
  \, + \, 
  e^{+\mu_\chi \delta_{\nu, 4}} \, \chi_x^\star \, 
  U_{x-\widehat{\nu}, \nu} \, \chi_{x-\widehat{\nu}}  \big] \!  \Big) \quad ,
\end{align}
where $M_\chi^2 = 8 + m_\chi^2$, and $\lambda_\chi$ is
the quartic coupling parameter for the field $\chi$. Also here we introduce a chemical
potential $\mu_\chi$, coupling to the temporal hopping terms. Note that compared to the
hopping terms in (\ref{phiaction}), for the field $\chi$ the link variables $U_{x,\nu}$
enter with an additional complex conjugation, implying that the fields $\phi$ and $\chi$
carry opposite charges.
Throughout this study we use an iso-spin like chemical potential by setting $\mu_\phi=\mu_\chi=\mu$.

The fact that for finite chemical potential $\mu_\phi$ and/or $\mu_\chi$, this theory has a complex sign
problem, implies that in the conventional representation the Boltzmann factor can no longer
be used as weight in a Monte Carlo simulation. This is why we here use a reformulation of the
theory in terms of \textit{dual variables} \cite{thisone}, which is briefly discussed in the next section.

\section{Dual representation of the partition sum}
\label{dualpartsec}
To avoid the sign problem of the conventional formulation as introduced
in the previous section, we reformulate the theory in terms of dual variables. In the dual formulation
the sign problem is gone and the real and positive weight factors occurring in the dual partition sum
can be used as weights in a Monte Carlo simulation. We will here just give the results of the dual 
reformulation and for a detailed treatment refer the reader to \cite{thisone}.

In a very compact notation, the partition sum in the dual formulation can be written as
\begin{eqnarray}
  Z =
  \left( \prod_{x, \nu<\tau} \sum_{p_{x,\nu\tau}=-\infty}^\infty \right) \!\!
  \left( \prod_{x,\nu} \sum_{l_{x,\nu}=-\infty}^\infty \sum_{l'_{x,\nu}=0}^\infty  
  \sum_{j_{x,\nu}=-\infty}^\infty \sum_{j'_{x,\nu}=0}^\infty \right) \!
  \mathcal{C}_g[p,l,j] \;
  \mathcal{C}_s[l] \;
  \mathcal{C}_s[j] \;
  \mathcal{W}_U[p] \;
  \mathcal{W}_\phi[l,l'] \;
  \mathcal{W}_\chi[j,j'] ,
  \nonumber
  \label{partsum}
\end{eqnarray}
where the dual, integer degrees of freedom are the plaquette variables $p_{x,\nu\tau}$, corresponding to the
original gauge degrees of freedom $U$, the link variables $l$ and $l'$ represent the original
d.o.f.\ of the field $\phi$, and the link variables $j$ and $j'$, correspond to $\chi$.

From
integrating out the original degrees of freedom we obtain constraints on admissible
configurations of the system, which in (\ref{partsum}) are denoted by the symbols $\mathcal{C}_g[p,l,j]$,
$\mathcal{C}_s[l]$ and $\mathcal{C}_s[j]$.
The constraint $\mathcal{C}_g[p,l,j]$ requires closed surfaces of
plaquette flux $p$, which have to be saturated by link fluxes $l$ and $j$ along open boundaries, while
the constraints $\mathcal{C}_s[l]$ and $\mathcal{C}_s[j]$ demand a conservation of $l$ and $j$ flux at
each lattice site. The primed link variables $l'$ and $j'$ are not subject to any constraint.

Admissible configurations come with a total weight $\mathcal{W}_U[p] \, \mathcal{W}_\phi[l,l'] \, \mathcal{W}_\chi[j,j']$,
where, as indicated by the notation, the total weight factorizes into a weight $\mathcal{W}_U[p]$ coming from the plaquettes,
another weight $\mathcal{W}_\phi[l,l']$ coming from the $l$ and $l'$ links, and a third weight $\mathcal{W}_\chi[j,j']$ 
which depends on the configuration of $j$ and $j'$ links. In contrast to the conventional form of this
theory, in the dual representation the weights are real and positive, even at finite chemical potentials $\mu_\phi$, $\mu_\chi$.
Using the dual representation allows one to perform a Monte Carlo simulation of the system at finite density.

\section{Observables}
\label{obssec}
In the dual like in the conventional representation, thermodynamic observables can be obtained as derivatives
of the logarithm of the partition sum with respect to the various couplings. The plaquette expectation value $\langle U \rangle$ and the corresponding susceptibility $\chi_U$, are given by
\begin{eqnarray}
\langle U \rangle =  \frac{1}{6 N_s^3 N_t} \, \frac{\partial}{\partial \beta} \, \ln Z 
\quad , \qquad
\chi_U  = \frac{1}{6 N_s^3 N_t} \, \frac{\partial^2}{\partial \beta^2} \, \ln Z 
\quad ,
%\label{eq:u1f2obsplaq}
\nonumber
\end{eqnarray}
i.e., they are obtained as derivatives with respect to the inverse coupling $\beta$. The field expectation values $\langle \phi^\star \phi \rangle$ and $\langle \chi^\star \chi \rangle$, together with the corresponding
susceptibilities $\chi_{\phi^\star \phi}$ and $\chi_{\chi^\star \chi}$, are
\begin{eqnarray}
  \langle \phi^\star \phi \rangle =  \frac{1}{V} \, \frac{\partial}{\partial M_\phi^2} \, \ln Z \; , \;
  \; \chi_{\phi^\star \phi}  = \frac{1}{V} \, \frac{\partial^2}{\partial (M_\phi^2)^2} \, \ln Z 
  \; , \; \; 
  \langle \chi^\star \chi \rangle =  \frac{1}{V} \, \frac{\partial}{\partial M_\chi^2} \, \ln Z \; , \; \;
  \; \chi_{\chi^\star \chi}  = \frac{1}{V} \, \frac{\partial^2}{\partial (M_\chi^2)^2} \, \ln Z
  \; ,
%  \label{eq:u1f2fieldobsdef}
\nonumber
\end{eqnarray}
and can be obtained as derivatives with respect to the squared mass parameters $M_\phi^2$ and $M_\chi^2$. The particle number densities $n_\phi$ and $n_\chi$, together with the associated susceptibilities
$\chi_{n_\phi}$ and $\chi_{n_\chi}$, are
\begin{equation}
  n_\phi = \frac{1}{V} \, \frac{\partial}{\partial \mu_\phi} \, \ln Z 
  \; , \; \;
  \chi_{n_\phi} = \frac{1}{V} \, \frac{\partial^2}{\partial \mu_\phi^2} \, \ln Z 
  \; , \; \;
  n_\chi = \frac{1}{V} \, \frac{\partial}{\partial \mu_\chi} \, \ln Z 
  \; , \; \; 
  \chi_{n_\chi} = \frac{1}{V} \, \frac{\partial^2}{\partial \mu_\chi^2} \, \ln Z 
  \; .
%   \label{eq:u1f2obsndef}
\nonumber
\end{equation}
 
To determine the $\mu=0$ mass spectrum of the theory we use the variational approach, where we diagonalize the connected correlation matrix $C_{ij}(\Delta n_t)$ by solving the eigenvalue equation
\begin{equation}
  C(\Delta n_t) \psi^{(l)} = \lambda^{(l)}(\Delta n_t) \psi^{(l)} \quad ,
\end{equation}
with the correlation matrix given element-wise by
\begin{equation}
  \label{eq:u1f2diffcorrmat}
  C_{ij}(\Delta n_t) = \langle \tilde{O}_i(\Delta n_t) \tilde{O}^\star_j(0) \rangle 
  - \langle 0 | \tilde{O}_i | 0 \rangle \langle 0 | \tilde{O}_j | 0 \rangle
  = \sum_l \langle 0 | \tilde{O}_i | l \rangle \langle l | \tilde{O}^\dagger_j | 0 \rangle 
  e^{-\Delta n_t E_l} \quad ,
\end{equation}
and the zero-momentum interpolators $\tilde{O}_i$ defined as
\begin{align}
  \label{eq:u1f2diffinterchoice}
  &\tilde{O}_1(n_t) = \frac{1}{V} \sum_{\vec{n}} \phi^\star(\vec{n},n_t) \, \phi(\vec{n},n_t)
  \quad , \qquad
  \tilde{O}_2(n_t) = \frac{1}{V} \sum_{\vec{n}} \phi(\vec{n},n_t) \, \chi(\vec{n},n_t) \quad , \\
  &\tilde{O}_3(n_t) = \frac{1}{V} \sum_{\vec{n}} \phi^\star(\vec{n},n_t) \, \chi^\star(\vec{n},n_t)
  \quad , \qquad
  \tilde{O}_4(n_t) = \frac{1}{V} \sum_{\vec{n}} \chi^\star(\vec{n},n_t) \, \chi(\vec{n},n_t) \quad .
\end{align}
Here the sums run over all lattice sites $\vec{n}$ and we used the unconventional combinations $\phi \chi$ and $\phi^\star \chi^\star$ in the interpolators because the action given in Section \ref{secconvaction} implies that the fields $\phi$ and $\chi$ are charge conjugate to each other.
It can be shown that the eigenvalues $\lambda^{(l)}$ are related to the masses $E_l$ of the physical states $\Psi^{(l)}$ by
\begin{equation}
  \label{eq:u1f2diffmasswhatisl}
  \lambda^{(l)}(\Delta n_t) \propto \exp{(-\Delta n_t E_l)} 
  \big(1 + \mathcal{O}(\exp{(-\Delta n_t \Delta E_l)}\big) \quad ,
\end{equation}
and further that the corresponding eigenvectors $\psi^{(l)}$ encode the contributions from the interpolators in the correlation matrix $C_{ij}$ to the physical states, 
\begin{equation}
  \Psi^{(l)} = \sum_i \psi_i^{(l)} \tilde{O}_i \quad .
\end{equation}
Note that the used interpolators correspond to two-particle bound states, since we probe the system in the confined phase and expect the lowest excitations to be of mesonic type.

\section{Results}
\begin{figure}
  \centering
  \includegraphics{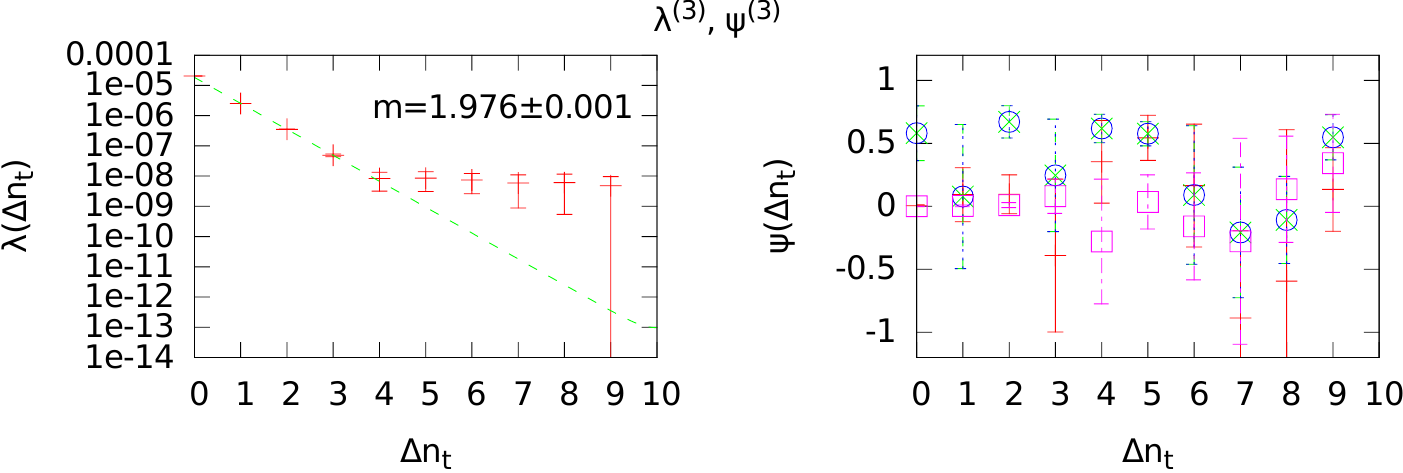}
  \caption{We show results for the relevant eigenvalue $\lambda^{(3)}$ and the corresponding eigenvector $\psi^{(3)}$ for the parameters $\beta=0.7$, $\lambda=1$, $M_\phi^2=5.3$, $M_\chi^2=5.7$ at $N_s^3 \times N_t = 16 ^3 \times 20$. In the l.h.s.\ plot we performed a fit to the data points at $\Delta n_t=1,2,3$.}
  \label{evmass}
\end{figure}
On the l.h.s.\ of Figure \ref{evmass} we show results for the eigenvalue $\lambda^{(3)}$ of the connected correlation matrix $\{C_{ij}\}$ given in (\ref{eq:u1f2diffcorrmat}), while on the r.h.s.\ we show the corresponding eigenvector $\psi^{(3)}$ as function of $\Delta n_t$. The used parameters were $\beta=0.7$, $\lambda=1$, $M_\phi^2=5.3$ and $M_\chi^2=5.7$ at a lattice size of $N_s^3 \times N_t = 16 ^3 \times 20$. We used $10^6$ sweeps for equilibration and performed $10^5$ measurements, seperated by $10$ decorrelation sweeps from each other. The eigenvalues $\lambda^{(4)}$, $\lambda^{(3)}$, $\lambda^{(2)}$, $\lambda^{(1)}$ are sorted in decreasing order and so the shown state $\lambda^{(3)}$ corresponds to the {\em second} lightest state in the physical spectrum. In the r.h.s.\ plot, red plusses encode the contribution from $\phi^\star \phi$ to the physical eigenstate $\Psi^{(3)}$, green crosses stand for $\phi \chi$, blue circles for $\phi^\star \chi^\star$ and magenta boxes for $\chi^\star \chi$. From the obtained results we conclude that the state $\lambda^{(3)}$ is the symmetric or anti-symmetric combination $(\phi \chi \pm \phi^\star \chi^\star) / \sqrt{2}$, associated with a mass of $m \approx 1.98$, which was obtained from a fit to $\lambda^{(3)}$ at $\Delta n_t=1,2,3$. We do not show the lightest state $\lambda^{(4)}$, because it turns out to be of the form $\phi^\star \phi$, which is a combination we do not expect to be excited by the introduced iso-spin like chemical potential $\mu_\phi=\mu_\chi=\mu$. Naively one expects that the condensation threshold is related to the mass of the lightest state $m$ the chemical potential couples to, by $2 \mu_c = m$, where in our case this state is one of the two combinations $(\phi \chi \pm \phi^\star \chi^\star) / \sqrt{2}$. Guided by \textit{Hund's rule} we tentatively make the assumption that the state with the higher multiplicity is the one with lower energy and so we expect the threshold of condensation to be connected to the mass of the triplet state $(\phi \chi + \phi^\star \chi^\star) / \sqrt{2}$, thus identifying $\lambda^{(3)}$ with the symmetric combination (consistent with the r.h.s.\ plot in Figure \ref{evmass}).

\begin{figure}
  \includegraphics{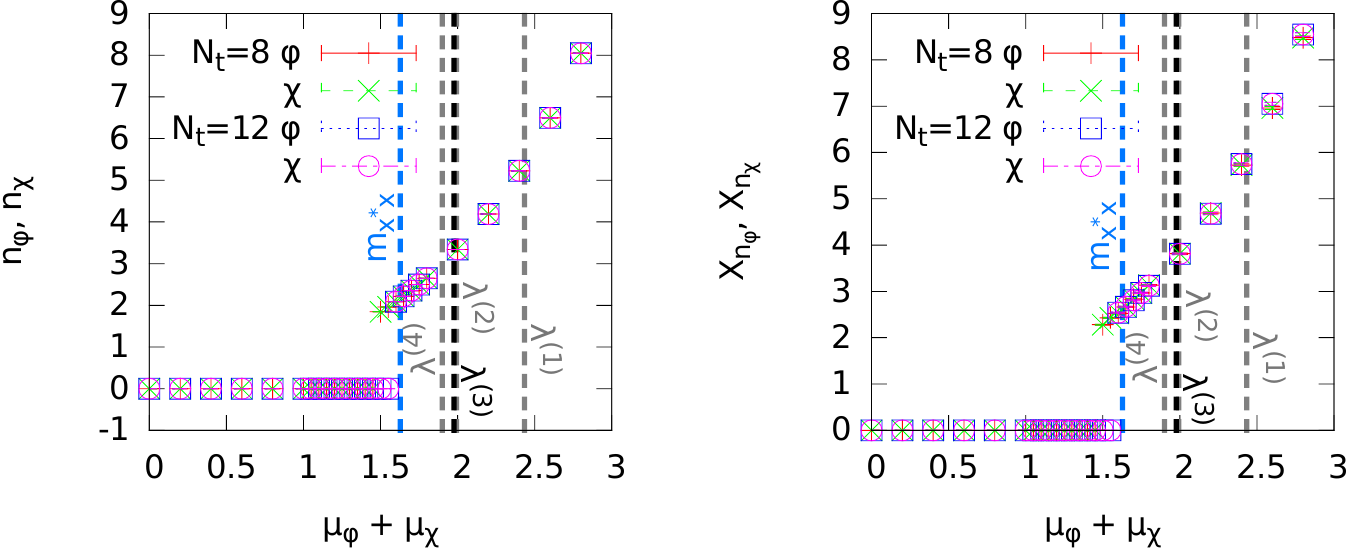}
  \caption{We show the particle densities $n_\phi$ and $n_\chi$ and the corresponding susceptibilities $\chi_{n_\phi}$ and $\chi_{n_\chi}$ as function of the chemical potential for $\beta=0.7$, $\lambda=1$, $M_\phi^2=5.3$, $M_\chi^2=5.7$ at $N_s^3 \times N_t = 8^3 \times 8,12$.}
  \label{figfin}
\end{figure}
In Figure \ref{figfin} we show the particle number densities $n_\phi$ and $n_\chi$ for the respective flavors and the corresponding susceptibilities as function of the chemical potential. The parameters were again set to $\beta=0.7$, $\lambda=1$, $M_\phi^2=5.3$ and $M_\chi^2=5.7$ at a spatial lattice volume of $N_s^3 = 8^3$, where we used $10^6$ equilibration steps and performed $10^5$ measurements seperated by $10$ steps. We observe a single threshold of condensation at $\mu_c \approx 1.6$ in all shown observables and also in the field expectation values $\langle |\phi|^2 \rangle$ and $\langle |\chi|^2 \rangle$ and the plaquette expectation value $\langle U \rangle$ which are not shown here because of limited space. The threshold in a run at higher temperature ($N_t=8$) is shifted towards smaller $\mu$ as is expected due to the higher thermal excitation of the system. The mass of the state $\lambda^{(3)}$ and also the other eigenvalues of the correlation matrix $C_{ij}$ are plotted as vertical lines, where it can be seen that the critical chemical potential does not match with the mass of $\lambda^{(3)}$, as was naively expected. 

We will suggest an explanation for the observed threshold in the following subsection.

\subsection{Relating the condensation threshold to the lowest excitation of the system}
We define the fields $x_x$ and $y_x$ at position $x$ as linear combinations of the fields $\phi_x$ and $\chi_x$ appearing in the original action given in Section \ref{secconvaction} 
\begin{equation}
  \label{eq:udiffdefxy}
  x_x \equiv \frac{\phi_x+\chi_x^\star}{\sqrt{2}}
  \quad , \qquad
  y_x \equiv \frac{\phi_x-\chi_x^\star}{\sqrt{2}} \quad .
\end{equation} 
In terms of the fields $x$ and $y$ we can reexpress the condensate which forms at the critical chemical potential and which we expect, as was argued before, to be of the form $\phi_x \chi_x + \phi_x^\star \chi_x^\star$ as
\begin{equation}
  \phi_x \chi_x + \phi_x^\star \chi_x^\star = |x_x|^2 - |y_x|^2 \quad .
\end{equation}
Since in the condensed phase the above expression will become large, we make the simplifying guess that at $\mu_c$ the modulus $|x_x|$ will also grow large, while $|y_x|$ will become very small. Then reexpressing the original action in terms of the fields $x$ and $y$ and ignoring contributions which are at least linear in $|y|$ due to $|y|$ being small, we obtain the reduced action $S_c[U,x]$:
\begin{align}
  \label{eq:u1diffredaction}
  \lim_{|y| \rightarrow 0} S[&U,x,y] \rightarrow S_c[U,x] =
  - \beta_c \, \sum_x \sum_{\sigma < \tau} \mbox{Re} \; U_{x,\sigma} U_{x+\widehat{\sigma}, \tau}
  U_{x+\widehat{\tau},\sigma}^\star U_{x,\tau}^\star
  \\
  &+\sum_x \!\Big( M_c^2 \, |x_x|^2  + \lambda_c |x_x|^4  -
  \sum_{\nu = 1}^4 \!
  \big[ x_x^{\star} \, U_{x,\nu} \, x_{x+\widehat{\nu}} 
  \, + \, 
  x_x^\star \,
  U_{x-\widehat{\nu}, \nu}^\star \, x_{x-\widehat{\nu}}  \big] \!  \Big) \quad .
  \nonumber
\end{align}
Where the reduced parameters are given by
\begin{equation}
  \label{eq:u1difforigparas}
  M_c^2 = \frac{M_\phi^2+M_\chi^2}{2} 
  \quad , \qquad
  \lambda_c = \frac{\lambda}{2} 
  \quad , \qquad
  \beta_c = \beta
  \quad .
\end{equation}
Note that since $\lambda_c = \frac{\lambda}{2}$ we expect the mass of the field $x$ to be smaller than the masses of the original fields $\phi$ and $\chi$.
So in the condensed phase the system is described by the reduced action $S_c[U,x]$ and we expect the lowest mesonic excitation of this reduced system to be related to the threshold of condensation shown in Figure \ref{figfin}: Thus we expect $2 \mu_c = m_{x^\star x}$, where the mass $m_{x^\star x}$ corresponds to the two-particle bound state formed by one $x$- and one anti-$x$ particle at zero chemical potential, with the dynamics of the field $x$ described by the reduced action $S_c[U,x]$ given in (\ref{eq:u1diffredaction}).

We extract the mass $m_{x^\star x}$ of simulations with the reduced action $S_c[U,x]$ at the reduced parameters
\begin{equation}
  M_c^2 = \frac{M_\phi^2+M_\chi^2}{2} = 5.5 
  \quad , \qquad
  \lambda_c = \frac{\lambda}{2} = 0.5
  \quad , \qquad
  \beta_c = \beta = 0.7
  \quad ,
\end{equation}
which are set according to (\ref{eq:u1difforigparas}) with respect to the original parameters we used in the finite chemical potential plots in Figure \ref{figfin}, so we can compare the obtained mass to the determined threshold. The result for the mass $m_{x^\star x}$ is shown in Figure \ref{figfin} as dashed vertical line labeled $m_{x^\star x}$, where it can be seen that the obtained mass does match the threshold very well. We take this as positive test for the explanation given above.

\section{Summary}
Within the studied model we suggested and verified an explanation for the non-trivial relation between the obtained $\mu_c$ threshold of condensation and the determined $\mu=0$ mass spectrum. In further studies it would be interesting to investigate the interaction between the excited mesons in the condensed phase to better understand the first-order nature of the condensation transition. It would also be appealing to introduce a third flavor into the theory to have a richer spectrum and as a consequence possibly encounter multiple condensation thresholds.

\end{document}